\documentclass{emulateapj}

\newcommand{\beq}{\begin{equation}}
\newcommand{\eeq}{\end{equation}}

\newcommand{\be}{\begin{equation}}
\newcommand{\ee}{\end{equation}}

\newcommand{\logLir}{{\rm log}[L_{\rm ir}/L_\odot]}
\newcommand{\logLirmin}{{\rm log}[L_{\rm ir}/L_\odot]_{\rm min}}
\newcommand{\zmax}{z_{\rm max}}

\begin{document}

\title{Correlations between Ultrahigh Energy Cosmic Rays and Infrared-Luminous Galaxies}

\author{
Andreas A. Berlind\altaffilmark{1},
Glennys R. Farrar\altaffilmark{2},
Ingyin Zaw\altaffilmark{2}
}
\altaffiltext{1}{Department of Physics and Astronomy, Vanderbilt University, Nashville, TN 37235, USA}
\altaffiltext{2}{Center for Cosmology and Particle Physics \& Department of Physics, New York University, New York, NY 10003, USA}


\begin{abstract}
We confirm the UHECR horizon established by the Pierre Auger Observatory using the heterogeneous Veron-Cetty Veron (VCV) catalog of AGNs, by performing a redshift-angle-IR luminosity scan using PSCz galaxies having infrared luminosity greater than $ 10^{10} ~L_\odot$.  The strongest correlation -- for $z<0.016$, $\psi=2.1^\circ$, and $L_{\rm ir} \geq 10^{10.5}~L_\odot$ -- arises in fewer than 0.3\% of scans with isotropic source directions.  When we apply a penalty for using the UHECR energy threshold that was tuned to maximize the correlation with VCV, the significance degrades to 1.1\%.  Since the PSCz catalog is complete and volume-limited for these parameters, this suggests that the UHECR horizon discovered by the Pierre Auger Observatory is not an artifact of the incompleteness and other idiosyncrasies of the VCV catalog.  The strength of the correlation between UHECRs and the nearby highest-IR-luminosity PSCz galaxies is stronger than in about $90$\% percent of trials with scrambled luminosity assignments for the PSCz galaxies.  If confirmed by future data, this result would indicate that the sources of UHECRs are more strongly associated with luminous IR galaxies than with ordinary, lower IR luminosity galaxies.  
\end{abstract}

\section{Introduction}
The Pierre Auger Observatory has reported \citep{augerScience07,augerlongAGN} a correlation between the arrival directions of the highest energy cosmic rays, and the positions of galaxies in the Veron-Cetty Veron ``Catalog of  Quasars and Active Galactic Nuclei" (12th Ed.) \citep{VCV} (VCV below).  Of the 27 cosmic rays above 57 EeV recorded by the Pierre Auger Observatory prior to Aug. 31, 2007, twenty are within 3.2$^\circ$ of a VCV galaxy with $z \leq 0.018$ (about 75 Mpc).  Restricting to $|b|>10^\circ$ to avoid regions in which galaxy catalogs are incomplete due to Galactic extinction, leaves 22 UHECRs of which 19 are within 3.2$^\circ$ of a VCV galaxy \citep{zbf10}.  The existence of a correlation between UHECRs and VCV galaxies clarifies the nature of ultrahigh energy cosmic rays by establishing that UHECRs are extragalactic in origin and that they have an energy-dependent horizon consistent with the GZK prediction.  However the VCV catalog is not complete, homogeneous, or pure, and its completeness drops with redshift \citep{VCV,zfg08,zbf10}.  Thus VCV is  not suitable for statistical analyses as would be required for establishing quantitative bounds on the contribution of various possible sources to the population of UHECRs, and its suitability for establishing the GZK horizon can be a source of concern.

Here, we study the correlation between UHECRs and infrared luminous galaxies with two aims.\footnote{Note that the terms ``luminous infrared galaxy'' (LIRG) and ``ultra-luminous infrared galaxy'' (ULIRG) commonly denote galaxies with $\logLir$ in the ranges 11-12 and $\geq 12$ respectively, but we are not using those conventional categories here.}
First, we use the {\em IRAS} point source catalog with redshifts (PSCz, \citealt{saunders_etal00}), which is homogeneous and complete to beyond the GZK distance, to replicate the Auger result of a UHECR horizon and correlation with extragalactic matter.   Second, we investigate whether the strength of a galaxy's correlation with UHECRs is related to its infrared luminosity, $L_{\rm ir}$.  In the absence of an active nucleus, infrared luminosity is a good proxy for star formation, and thus galaxies with strong IR emission should have an enhanced concentration of massive stars, whose lifetimes are short on the time scale of star formation episodes.  A number of proposed UHECR sources are associated with the death of massive stars or birth of their progeny, e.g., collapsars, GRBs, or magnetars, so the probability of such events occurring in a given galaxy could increase roughly in proportion to $L_{\rm ir}$ of the galaxy.  These considerations motivate a study of the correlation between UHECRs and high $L_{\rm ir}$ galaxies.   Of course, some high $L_{\rm ir}$ galaxies host AGNs, so discovering that correlations are enhanced with a high $L_{\rm ir}$ sample does not by itself demonstrate a non-AGN source population and further investigation is needed.  In particular, low-redshift GRBs occur primarily in relatively rare {\em low metallicity} star-forming regions \citep{stanek06}, especially found in dwarf galaxies, so a UHECR correlation with high luminosity IR galaxies would not necessarily be indicative of GRB sources.  The correlation between AGASA UHECRs and PSCz galaxies was investigated by \citet{LIRG-AGASA...Giller...02} and the clustering of Auger UHECRs was compared to that of PSCz by \citet{cuoco08}.

We use the technique employed by Auger, of scanning over the maximum angular separation, $\psi$, and over the maximum redshift of the galaxies (also see \citealt{tinyakov01}).  We also scan on the threshold infrared luminosity of galaxies.  Since only events with $E>57$~EeV have been published, we do not do a full scan on energy, but we do a partial scan in one direction (toward higher energy) in order to estimate the penalty associated with adopting the 57~EeV threshold.  To assess the significance of the correlation compared to isotropy, as relevant to establishing the existence of a UHECR horizon, we perform the same scans on mock isotropic UHECR datasets.  To test the significance of a possible correlation with infrared luminosity, we compare to the results of scans using mock catalogs obtained by scrambling the luminosities of the PSCz galaxies.
 
\section{ Scan Analysis with Infrared Luminous Galaxies}

\begin{figure}[t]
\epsscale{1.1}
\plotone{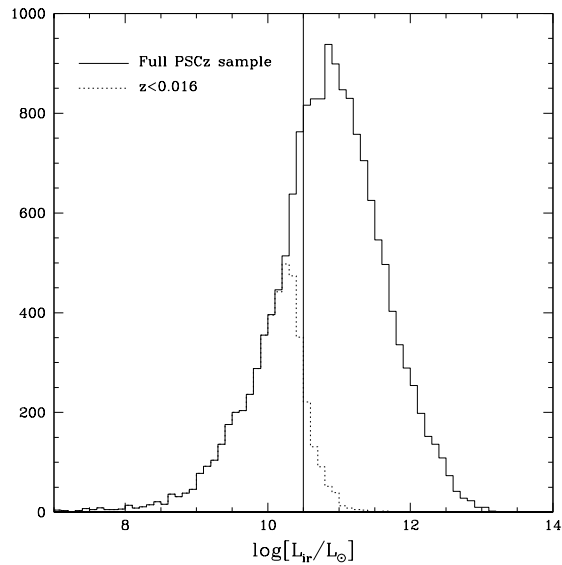}
\caption{
Luminosity distribution for the full PSCz catalog (solid) and within $z<0.016$ (dotted). The vertical line at 10.5 indicates the luminosity threshold for maximally significant UHECR correlations.  Note that there are very few luminous PSCz galaxies at low redshift.}
\label{Ldist}
\end{figure}

{\em IRAS} is an all-sky survey of infrared galaxies, and the PSCz catalog provides the redshifts of {\em IRAS} point sources.  
\citet{SandersandMirabel1996} give the relationship between the {\em IRAS} observed fluxes at 12, 25, 60 and 100$\mu$m in W/m$^{2}$ ($f_{12}$, etc.), and the total 8-1000$\mu$m infrared luminosity:
\begin{equation}
\label{LIR}
L_{\rm ir}   = 4 \pi D_L^2 \,1.8 \times 10^{-14} [13.48 f_{12} + 5.16 f_{25} + 2.58 f_{60} + f_{100}] {\rm W/m^2},
\end{equation}
where $D_L$ is the luminosity distance.  Figure~\ref{Ldist} shows the distribution of $\logLir$ in the entire PSCz catalog (solid), and, for later reference, for $z<0.016$ (dotted).  We implement the PSCz mask by a simple Galactic latitude cut $|b| \geq 10^\circ$, the recommended conservative cut to avoid poorly sampled regions near the Galactic plane \citep{saunders_etal00}.

We perform a scan on the angular separation, maximum redshift and minimum infrared luminosity of PSCz galaxies and UHECRs with $|b|>10^\circ$, covering the ranges $\psi=0.1^\circ-5^\circ$, $\zmax=0-0.024$, and $\logLirmin=10-11$, in steps of $\Delta\psi=0.1^\circ$, $\Delta z=0.001$, and $\Delta\logLirmin=0.1$.  
For each combination of $\zmax$, $\psi$, and $\logLirmin$ we find the number of correlated UHECRs, $k_{\rm corr}$, and compute the associated probability measure $P$.  $P$ is the cumulative binomial probability to find $k_{\rm corr}$ correlated events out of $N_{\rm CR}$ cosmic rays, given the fractional sky coverage $p$ of PSCz galaxies within the given redshift and luminosity bounds.  $p$ is the exposure-weighted projection of disks of radius $\psi$ around each galaxy in the sample, normalized to the total Auger exposure; it must be recomputed for each bin in $\zmax$, $\psi$, and $\logLirmin$.  The minimum value of $P$ in the scan analysis is found for $\zmax=0.016,~ \logLirmin=10.5$, and $\psi=2.1^\circ$.  With these parameters, there are 632 PSCz galaxies, $p=0.1611$, and 13 out of 22 UHECRs correlate.  The associated probability measure is $P=5.7\times 10^{-6}$.  The probability of finding as good or better correlation by chance from an isotropic source distribution is $(2.68 \pm 0.07) \times 10^{-3}$.  This is measured by creating $5 \times 10^5$ catalogs of 22 random UHECR directions  with $|b|>10^\circ$ weighted by the Auger exposure, and using these to perform the scans with the PSCz galaxy sample rather than real Auger data.

\begin{figure}[h]
\epsscale{1.1}
\plotone{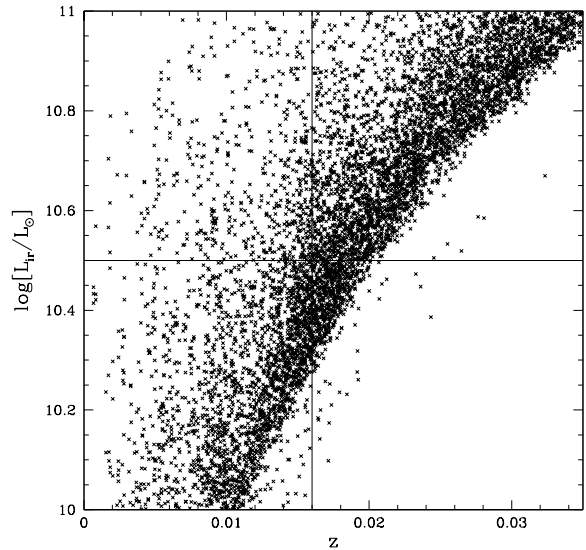}
\caption{
Infrared luminosity vs. redshift for PSCz galaxies.  The vertical and horizontal lines at indicate the values of $\logLirmin$ and $\zmax$ for which UHECR correlations are maximum.}
\label{LzPSCz}
\end{figure}

The choice of energy cut ($E>57$~EeV) made by Auger in releasing the data was tuned to maximize the correlation of UHECRs with VCV AGN.  While this analysis is independent of the VCV analysis performed by Auger, VCV and PSCz galaxies both trace the same underlying matter density field.  It is thus reasonable to worry that the choice of energy cut will bias our results.  In order to estimate a penalty for this, we include a scan in energy threshold in one direction, toward higher energy.  We find that the minimum value of $P$ occurs for the original energy cut of 57~EeV and the same values of $\psi$, $\zmax$, and $\logLirmin$.  We then use $5 \times 10^4$ catalogs of 22 random UHECR directions in new scans that include the energy scan in one direction.  We find that the probability of finding as good or better correlation by chance from an isotropic source distribution is $(1.10 \pm 0.05)\times 10^{-2}$.  In other words, the significance of our result degrades from 0.3\% to 1.1\% when we include the energy penalty.

As can be seen from Figure~\ref{LzPSCz}, the PSCz catalog is complete to redshift 0.024 for $\logLir \geq 10.6$, to 0.02 for $\logLir \geq 10.5$, and for $z \leq 0.016$ it is complete down to $\logLir < 10.3$.  The scan minimum ($\logLirmin=10.5$, $\zmax=0.016$) is thus comfortably inside the domain of PSCz completeness and is thus established with a volume-limited sample.  Although the whole scan range includes regions for which the PSCz catalog is not complete, the simulations measuring the chance probability for isotropically distributed UHECRs to give as low or lower a $P$ value have the same incompleteness, and therefore give a reliable estimate of the significance.  Therefore, the analysis presented here suggests that the arrival directions of UHECRs correlate with nearby extragalactic objects and establishes the UHECR horizon.
 
\section{Infrared Luminosity and UHECR Correlation}

\begin{figure}[t]
\epsscale{1.1}
\plotone{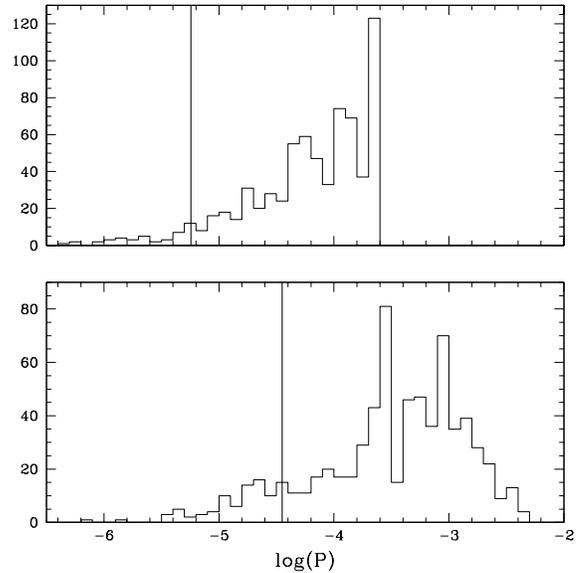}
\caption{
Upper panel: histogram of $P$ values obtained in 700 scans of UHECRs and PSCz galaxies, but for scrambled values of $\logLir$; the $P$ value for the true $\logLir$ is shown as a vertical line.  Lower panel: same, but with 2 UHECRs correlating with VCV galaxies removed.}
\label{LIRscramble}
\end{figure}

One next wants to know if this correlation of UHECRs with luminous PSCz galaxies is stronger than if the galaxies were selected from the PSCz catalog without regard to their IR luminosity.  To find out, we follow the same scan procedure as above, but scrambling the values of $\logLir$ first.  The $P$ values obtained in 700 such trials are shown in the upper panel of Figure~\ref{LIRscramble}.  Six percent of the cases have a $P$ value as low as in the real scan using the highest luminosity IR galaxies.  

Two of the VCV galaxies that correlate with pre-prescription UHECRs are also in the PSCz catalog.  Thus the formulation of the Auger prescription might indirectly bias the present study of UHECR-LIRG correlations.  We can remove any such bias in estimating the chance probability, at the price of obtaining only a bound on the significance, by removing those two UHECRs from the UHECR data set.  Since that explicitly eliminates some correlations, we find an upper bound on the {\em a priori} chance probability.   The lower panel of Figure~\ref{LIRscramble} shows the distribution of $P$ values for the reduced UHECR data set and scrambled IR luminosity values;  the vertical line is the $P$ value found using the reduced UHECR data set and scanning on the true IR luminosity.  The observed $P$ value is lower than in the scrambled data sets in 11.7\% of the cases, allowing us to bracket the chance probability for the UHECR-LIRG correlation not being $L_{\rm ir}$ dependent to be in the range 6-12\%.  This result suggests that UHECRs are more strongly correlated with highly luminous PSCz galaxies than lower luminosity galaxies, but it is not statistically significant and remains to be confirmed with future data.

Table~1 gives the name, Right Ascension and declination of the PSCz galaxies that correlate with UHECRs for the scan parameters that minimize $P$ ($z \leq 0.016$, $\psi \leq 2.1^\circ$ and $\logLir \geq 10.5$), followed by the year, Julian day, and energy of the correlated UHECR.  This is followed by the angular separation between galaxy and UHECR, and the redshift and derived infrared luminosity of the galaxy.  In some cases, PSCz reports multiple values of $L_{\rm ir}$; in such cases these are listed separated by a comma. Thirteen distinct PSCz galaxies correlate with 13 different UHECRs within $2.1^\circ$, with  two UHECRs correlating with more than one PSCz galaxy and two PSCz galaxies correlating with more than one UHECR: 2MASX J17544125-6054404 ({\em IRAS} 17501-6054) and ESO 270- G 007 ({\em IRAS} 13203-4317).  Of course, a given cosmic ray can have only one source, and the presence of two source candidates within the scanning prescription is a useful reminder of the limitations of the scanning method.  

It must be emphasized that a correlation in the scan procedure does not imply a correct identification of the CR source: the true source for that cosmic ray may not even be in the catalog or may be beyond the $\zmax$ which minimizes $P$.  Only a significant excess of correlations compared to chance, for the ensemble of correlations, is evidence for a concentration of sources in the catalog.  The Generalized Maximum Likelihood method \citep{fGML08}, which allows the GZK redshift information to be imposed on an event-by-event basis, is distinctly superior to the scan method in this regard (Busca and Farrar, 2009, in preparation) but we have adopted the scan method for the present work, to allow direct comparison to the Auger analysis with the VCV catalog.  Use of the new Generalized Maximum Likelihood method should significantly improve the sensitivity of future correlation studies (Zaw, Busca, Farrar, 2010, in preparation).  Since PSCz is complete for the redshift and $L_{\rm ir}$ range of interest, and outside its mask the PSCz catalog does not suffer from the inhomogeneity issues plaguing the VCV catalog, detailed statistical studies to gain further insight into the origin of the correlation will be possible.  

Both AGN activity and star formation lead to a high $L_{\rm ir}$ after reprocessing by dust, and indeed 5 of the 13 correlated PSCz galaxies have optically identified active nuclei by the criteria described in \citet{zfg08}.  This is a much larger portion than for randomly chosen high IR luminosity galaxies, since studies show that 4\% of galaxies with $\logLir$ in the range $10-11$ are Seyferts, i.e. confirmed optical AGNs \citep{SandersandMirabel1996}.  Only about half of all active nuclei can be identified in the optical \citep{reviglioHelfand06}, so an important question is whether any of the 7 correlated PSCz galaxies that do not appear in VCV have active nuclei as well.  Upcoming X-ray observations with Chandra of these galaxies will help settle this question.   For the moment, we cannot determine whether the UHECR correlation with IR luminous galaxies may result from AGNs they contain, or may represent evidence for a  second class of sources.

\section{ Conclusions}

The scan analysis performed here shows a significant correlation between the published Auger cosmic rays and nearby moderately-to-highly luminous infrared galaxies.  This provides a valuable independent corroboration of the UHECR horizon with a homogeneous, volume-limited catalog.  The correlation is maximized for $z \leq 0.016$, $\psi = 2.1^\circ$ and $\logLirmin=10.5$.   The probability that isotropic sources would show such a high degree of correlation by chance is 0.3\%.  The PSCz catalog is complete to $z = 0.024$ for $\logLir \geq 10.5$, thus we establish the UHECR horizon with a volume-limited catalog, removing the concern that the horizon found in the Auger analysis with VCV galaxies could be an artifact of VCV's increasing incompleteness with distance \citep{zbf10}.  

The angular separation maximizing the correlation in the PSCz luminosity scan, $\psi = 2.1^\circ$, is lower than the $\psi = 3.2^\circ$ optimizing the VCV scan.  The relationship between $\psi$ and the spectrum of magnetic deflections is subtle and depends on the source sample size and completeness, since $\psi$ is determined by minimizing $P$.   Nonetheless, this difference may be evidence that the incompleteness and inhomogeneity of VCV exaggerates the angular separation between source and UHECR arrival direction, and thus one must be cautious about inferring the typical magnetic deflection at this time. 

The correlation observed between the published Auger UHECRs and highly luminous PSCz galaxies is stronger than between UHECRs and randomly chosen PSCz galaxies within the same redshift and angular region, in 6-12\% of the trials.  If confirmed by future data, such a correlation between UHECRs and high-IR-luminosity galaxies would show that luminous IR galaxies have an increased abundance of sources of UHECRs compared to randomly chosen galaxies in the PSCz catalog.  It does not, however, resolve the question of whether UHECRs are produced by AGNs or some other mechanism, perhaps associated with the demise of stars as in GRBs or magnetars, since many of the correlated infrared galaxies are known to contain AGNs.  A program of Chandra observations is underway to find out which of the other correlated high-IR-luminosity galaxies may also have active nuclei.  

We note that new Auger data (since the published 2007 Science results) show a weaker correlation with the VCV AGN \citep{hand10}, highlighting that the published UHECR dataset may be too small to place robust constraints on cross-correlations with extragalactic sources.  It is thus especially important to repeat the present analysis with a new and larger dataset.

We acknowledge the essential role of the Pierre Auger Collaboration in this work, which is based on data obtained and published by the  Pierre Auger Observatory.  Furthermore, GRF and IZ acknowledge their membership in the Pierre Auger Collaboration and thank their colleagues for their participation in and contribution to this research.  We also thank J. Moustakas and A. Soderberg for helpful input, as well as the anonymous referee for comments that led to improvement of the paper.  This research has been supported in part by NSF-PHY-0701451.
\onecolumngrid

\onecolumngrid

\begin{deluxetable*}{cccccccccc}
\tablecolumns{10}
\tabletypesize{\scriptsize} 
\tablewidth{0pt}  
\tablecaption{\label{t:ass}
Properties of correlating PSCz galaxies
}
\tablehead{
IR Galaxy &  RA & Dec & AugerYr, Day &  E &  	 r  &     z  &  $L_{\rm ir}$  & AGN? & VCV Name\\ 
    &  (J2000)   & (J2000)&           & (EeV) & (deg) &        & ($L_{\odot}$) &
}
\startdata 
NGC 1346 & 52.55312 & -5.54282 & 2006, 296 & 69 & 1.04 & 0.013 & 10.52 & S1 & SDSS J03302-0532 \\
NGC 1204 & 46.16654 & -12.34082 & 2007, 145 & 78 & 1.56 & 0.014 & 10.91 & H2 & NGC 1204 \\
ESO 565-G006 & 141.80370 & -19.53184 &2007, 84 & 64 & 1.83 & 0.016 & 10.81 &  & \\
NGC 424 & 17.86485 & -38.08423 & 2005, 54 & 84 & 0.37 & 0.012 & 10.62 & S1.5 & NGC 424 \\
NGC 7130 & 327.08014 & -34.95164 & 2007, 193 & 90 & 2.00 & 0.016 & 11.35 & S2 & IC 5135 \\
NGC 4945 & 196.36290 & -49.46769 & 2005, 81 & 58 & 2.02 & 0.002 & 10.79,11.03  & S2 & NGC 4945 \\
ESO 270-G007 & 200.81441 & -43.54422 & 2006, 299 & 69 & 1.75 & 0.013 & 10.55 &  & \\
             &            &            & 2007, 69  & 70 & 0.49 &       &         &  & \\
NGC 5128 (Cen A)& 201.36003 & -43.01828 & 2007, 69 & 70 & 0.91 & 0.002 & 10.52,10.62 & BLLac & NGC 5128 \\
IC 4523 & 226.29191 & -43.50967 & 2004, 343 & 63 & 1.46 & 0.016 & 10.57 &  & \\
2MASX J17544125-605440 & 268.67473 & -60.91092 & 2004, 339 & 83 & 0.14 & 0.016 & 10.52 &  & \\
                        &            &            & 2006, 55 & 59  & 0.54 &       &         &    & \\
NGC 7648 & 350.97771 &  9.66911 & 2006, 185 & 83 & 0.93 & 0.012 &  10.55 &  &  \\
IC 5179 & 334.03740 & -36.84463 & 2005, 295 & 57 & 1.68 & 0.011 & 11.16 &  & \\
IC 5186 & 334.69389 & -36.80152 & 2005, 295 & 57 & 2.05 & 0.016 & 10.65 &  & \\
\enddata 
\tablecomments{
This table lists key properties of the {\em IRAS} PSCz galaxies with $|b|>10^\circ$ that are within 2.1$^\circ$ of an Auger UHECR. The arrival year and Julian day, and the energy of the correlated UHECR are given for each galaxy, followed by the separation between the IR galaxy and UHECR in degrees, and the redshift and derived infrared luminosity of the galaxy; when multiple values of $L_{\rm ir}$ are given in PSCz, all are listed, separated by a comma.  For galaxies appearing in the VCV catalog, the ZFG classification and VCV name for the galaxy is also given.
}
\end{deluxetable*}

\bibliographystyle{apj}
\bibliography{mnem,CR,AGN,uhecr,GRB-IRgals}

\end{document}